\documentclass[prb,twocolumn,english,superscriptaddress,floatfix,longbibliography]{revtex4-1}

\usepackage{graphicx}
\usepackage{amsmath}  
\usepackage{epigraph}
\usepackage[dvipsnames]{xcolor}
\usepackage{float}

\def\dto{Dy$_2$Ti$_2$O$_7$}

\def\latent{$\mathcal{L}_Q$\ }
\def\Sexp{$S^{\rm exp}(Q)$}
\def\Ssim{$S^{\rm sim}(Q)$}
\begin{document}

 
\title{Integration of Machine Learning with Neutron Scattering: Hamiltonian Tuning in Spin Ice with Pressure}

\author{A. M. Samarakoon}
\address{Neutron Sciences Division, Oak Ridge National Laboratory, Oak Ridge, TN 37831, USA.}
\address{Shull Wollan Center - A Joint Institute for Neutron Sciences, Oak Ridge National Laboratory, TN 37831. USA}
\address{Materials Science Division, Argonne National Laboratory, Argonne, IL, USA}

\author{D. Alan Tennant}
\address{Neutron Sciences Division, Oak Ridge National Laboratory, Oak Ridge, TN 37831, USA.}
\address{Shull Wollan Center - A Joint Institute for Neutron Sciences, Oak Ridge National Laboratory, TN 37831. USA}
	
\author{Feng Ye}
\address{Neutron Sciences Division, Oak Ridge National Laboratory, Oak Ridge, TN 37831, USA.}
	
\author{Qiang Zhang}
\address{Neutron Sciences Division, Oak Ridge National Laboratory, Oak Ridge, TN 37831, USA.}
	
\author{S. A. Grigera}
\address{Instituto de F\'{\i}sica de L\'{\i}quidos y Sistemas Biol\'ogicos, UNLP-CONICET, La Plata, Argentina}

\date{\today}

\begin{abstract}

Quantum materials research requires co-design of theory with experiments and involves demanding simulations and the analysis of vast quantities of data, usually including pattern recognition and clustering.  Artificial intelligence is a natural route to optimise these processes and bring theory and experiments together.  Here we propose a scheme that integrates machine learning with high-performance simulations and scattering measurements, covering the pipeline of typical neutron  experiments. Our approach uses nonlinear autoencoders trained on realistic simulations along with a fast surrogate for the calculation of scattering in the form of a generative model.  We demonstrate this approach in a highly frustrated magnet, Dy$_2$Ti$_2$O$_7$, using machine learning predictions to guide the neutron scattering experiment under hydrostatic pressure, extract material parameters and construct a phase diagram. Our scheme provides a comprehensive set of capabilities that allows direct integration of theory along with automated data processing and provides on a rapid timescale direct insight into a challenging condensed matter system.

\end{abstract}

\maketitle

\section*{}
Artificial Intelligence holds the promise of profound and far reaching impact on experimental science by integrating theory and experiment in new ways \cite{hey2020}. Neutron scattering on quantum materials is an area where much progress can be expected \cite{hey2020,Chen_2021,Doucet_2021} which would impact co-design of theory and experiment as well as materials discovery and optimization. To achieve this requires the integration of simulations, data treatment and analysis, and theoretical interpretation \cite{Doucet_2021}. Data science, and in particular machine learning (ML), have been proposed as ways to integrate scattering experiments with demanding state-of-the art simulations \cite{Chen_2021, Samarakoon:2020aa}, however effective schemes to do this remain to be demonstrated. Here, we deploy ML across the experimental pipeline, closely integrating theory and experiment in a way that could be used more widely for materials research. We apply it to a highly frustrated magnet that provides challenges representative of current quantum materials.

Traditionally, experiment planning, data treatment, and analysis have taken major efforts involving months of detailed work \cite{Tennant-2021,Doucet_2021,Chen_2021}. They have relied on often crude analytic approximations due to the difficulty in matching time consuming and highly specialized simulations with experiment. Recently we have shown that machine learning, and in particular the application of Non-Linear Autoencoders (NLAEs) can be used to create automated capabilities for Hamiltonian extraction from diffuse neutron scattering data \cite{Samarakoon:2020aa,Tennant-2021}. Further, this approach was demonstrated to provide robust parameter optimization, automated denoising and data treatment, as well as phase diagram mapping and categorization.

Here we show how a complete integration can be achieved and present a scheme that integrates the experiments with theory and modelling on the experiment timescales.  The approach in Ref. \onlinecite{Samarakoon:2020aa} is augmented with generative models to provide fast surrogates for expensive materials' simulations. This facilitates analysis to be conducted {\em during} experiments with the results feeding back into choices made by the experimenter. We demonstrate the key elements of ML that enable neutron experiments on the highly frustrated magnet Dy$_2$Ti$_2$O$_7$ \cite{Samarakoon:2020aa,Tennant-2021}. This material shows complex physical behavior that requires sophisticated simulations to understand and thus stands as an ideal test case. 
While our approach can be more closely integrated into experiment by connecting to the data collection and experimental steering, the capabilities are not yet in place to do this at the Spallation Neutron Source, Oak Ridge National Laboratory, and a number of steps are still carried out manually. However, the study here provides a proof-of-principle for deeper integration of machine learning into neutron scattering pipelines as well as allowing us to understand a complex condensed matter system on a rapid timescale. As such it provides a schema for the co-design of materials and theory more directly with experiment.

%
\begin{figure*}
\centering\includegraphics[width=0.99\textwidth]{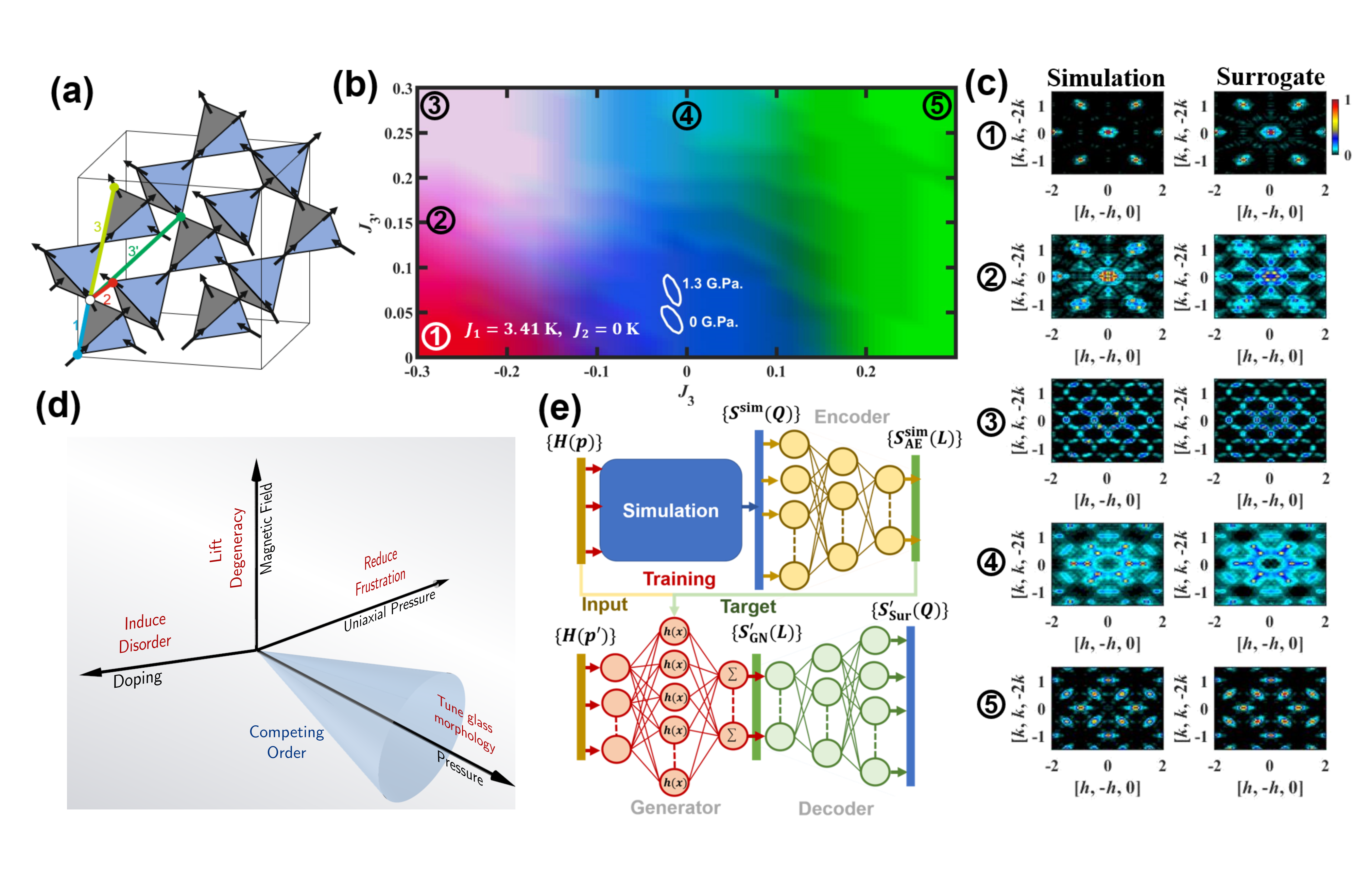}
	\caption{
	{\em Structure of the magnetic system, phase map and generative model.} 
	(a) The magnetic moments located on Dy ions are constrained by crystal field interactions to point in or out of the tetrahedra. They form a corner sharing pyrochlore lattice. Nearest neighbors (1), next-nearest-neighbors (2) and two inequivalent next-next-nearest neighbors (3 and 3') interactions are shown as thick colored lines. (b) A predicted map of magnetic orderings for varying $J_3$ and $J_{3'}$ with the remaining Hamiltonian parameters $J_1$, $J_2$, and $D$ fixed to 3.41 K. 0 K and 1.3224 K respectively. The latent space coordinates, $S(L)$ predicted by the Generator network (GN) have been converted to an RGB color code. A region with uniform color is expected to be structurally the same, and continuous color changes correspond to either crossovers or continuous transitions. (c) Comparison between the high-symmetry-plane slices of simulated and surrogate-predicted $S(Q)$ data at multiple places in parameter space as indexed on panel (b). (d) Schematic diagram of the experimental parameters that can be used to tune the properties of frustrated materials and their associated effects. (e) Schematic design of the surrogate model used to predict $S(Q)$ for a given set of model parameters. The surrogate comprises a radial basis network, the Generator, mapping parameter space to latent space and a Decoder to reconstruct $S(Q)$ from latent space representations.}
\label{fig01_genModel}
\end{figure*}

%
\begin{figure*}
\centering\includegraphics[width=0.99\textwidth]{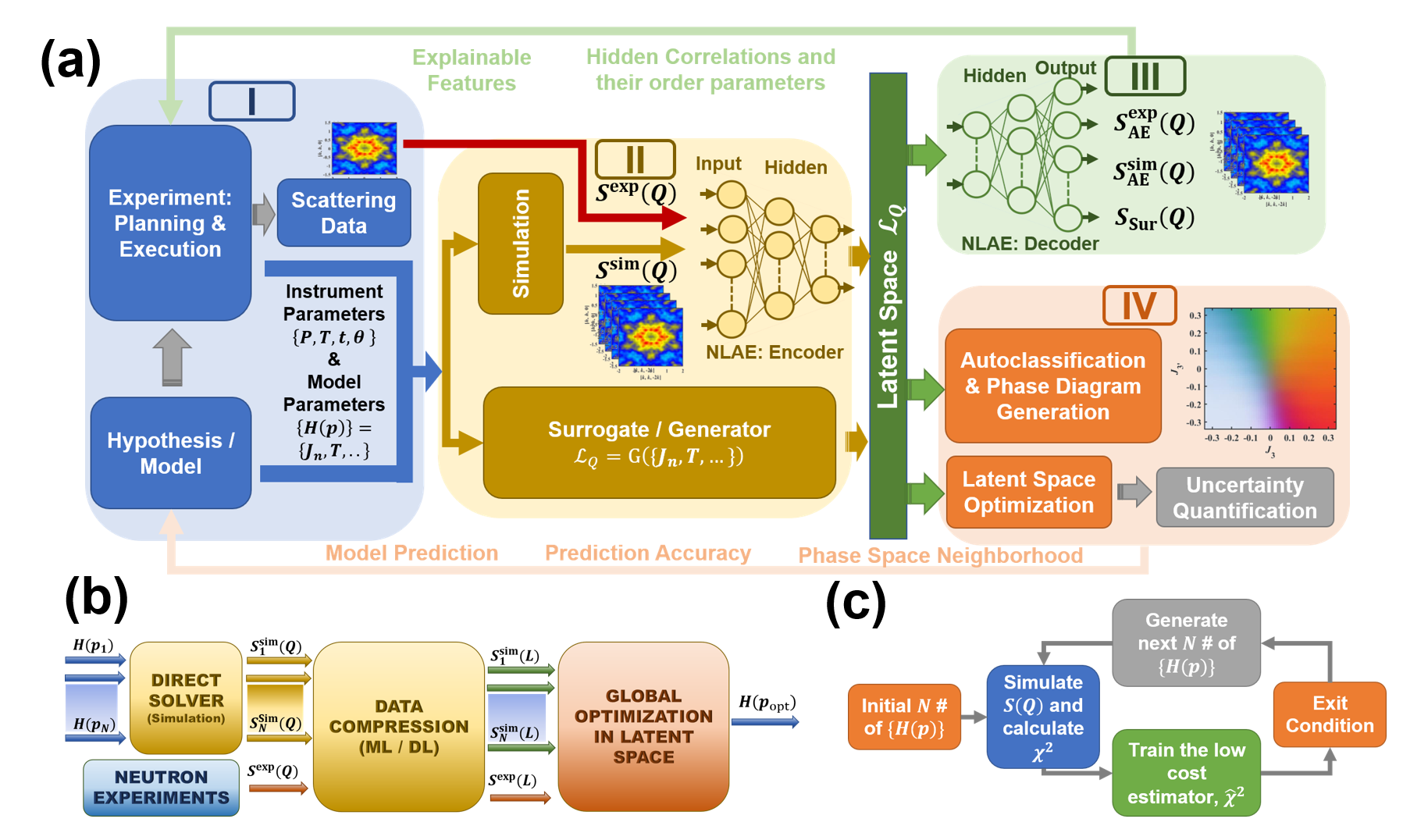}
\caption{
{\em Schematic overview of machine-learning integration into the direct and inverse scattering problem} (a) 
The ML workflow used here to drive the scattering experiment with autonomous data analysis and feeding back vital information.
The workflow is split into four main sections: (I) scattering experiment design and optimization; (II) parameter space exploration and information compression; (III) structure or property predictions; and (IV) parameter space predictions. Section II links to both III and IV via latent space, \latent, a compressed version of the large pixel space. The latent space representations, $S(L)$, are used in surrogates that bypass expensive calculations. These surrogates are used for exhaustive searches of parameter space, identifying phases and phase transitions, and predicting optimal regions for experimental study. More simulations are done iteratively in the areas of interest, and the surrogates are trained accordingly to improve their prediction accuracy. (b) A schematic representation of the proposed scheme to address the inverse scattering problem for neutron experiments. Global optimization is proposed to be performed, followed by a data compression step that transforms input $S(Q)$ into $S(L)$. (c) The detailed workflow of the global optimization algorithm used in this work  to determine the region in parameter space where the model best reproduces experiments.}
\label{fig02_MLoverview}
\end{figure*}

\subsection*{Case study: the complex magnetic properties of Dy$_2$Ti$_2$O$_7$}

We will showcase the ML aided approach to a neutron scattering experiment by studying the behaviour under hydrostatic pressure of Dy$_2$Ti$_2$O$_7$, a notable example of a   magnetically frustrated material. 

In geometrically frustrated materials, the dominant pairwise interactions cannot be simultaneously minimized due to constraints dictated by the arrangement of spins on the lattice. Intricate correlations as a result of the mutually struggling ordering tendencies become manifest in the ground states.  Real materials are rich systems, with multiple magnetic interactions covering a broad range of energies and length-scales. The neutralisation of the dominant forces leave the ground open for minor players to determine the outcome.  Frustration can thus be an avenue towards subtler, more exotic types of order at low temperatures, such as exponentially degenerate ground states similar to water ice \cite{bramwell2001spin}, fractionalized magnetic excitations \cite{castelnovo2008magnetic}, gigantic anomalous Hall effect \cite{taguchi2001spin}, spin-glasses and spin-liquid phases \cite{gardner2010magnetic}.

Spin-ices can be described as classical ferromagnetic Ising spins on the cubic pyrochlore lattice (see Fig.~\ref{fig01_genModel} a).  The system is expected to remain in an exponentially degenerate disordered state: a three dimensional spin-liquid with an emergent gauge field and fractionalised excitations \cite{castelnovo2012spin}. In real materials, such as Dy$_2$Ti$_2$O$_7$ (DTO) and Ho$_2$Ti$_2$O$_7$ (HTO), the situation is more complicated and several magnetic interactions are necessary to account for experimental observations \cite{melko2004monte, yavors2008dy}.  High-order dipolar terms are expected to order the system at very low temperatures \cite{melko2004monte}, but experiments show no long-range order down to the lowest temperatures where equilibrium properties are accessible, around 0.6 K, below which the system freezes \cite{snyder2001spin, snyder2004low}.

The material of our choice, Dy$_2$Ti$_2$O$_7$, is perhaps the cleanest spin-ice material, and as such it has been heavily studied. The system is under a delicate balance of interactions. Fits to experimental data result in a complex empirical Hamiltonian, with a nearest neighbour, $J_1$, next nearest neighbour $J_2$, and two inequivalent third nearest neighbour exchange interactions, $J_3$, $J_3'$,  (see Fig.~\ref{fig01_genModel} a), and a magnetic dipolar interaction term, $D$\cite{henelius2016refrustration,borzi2016intermediate}.
Previously we have demonstrated the application of non-linear autoencoders (NLAEs) to analyze neutron diffuse scattering, and used it to autonomously analyze and interpret the ambient pressure measurements on DTO \cite{Samarakoon:2020aa,Tennant-2021}. NLAEs were demonstrated to fulfill key tasks including denoising of data; treatment for background and experimental artifacts; as well as optimization of model parameters, and construction of a high dimensional phase diagram of the different ground states in the neighboring region of parameter space.
As expected for a frustrated system, even within a restricted region of parameter space  there is an abundance of competing phases. Experimentally, the control of external parameters such as pressure, both uniaxial\cite{edberg2019dipolar, edberg2020effects} and hydrostatic \cite{mirebeau2002pressure,mirebeau2004}, doping \cite{zhou2011high, borzi2013charge}, and magnetic field can be used as a powerful tool to search for new ordered states in frustrated systems (see Fig. \ref{fig01_genModel}(d)). This opens a vast multi-dimensional space to be explored and one where ML enabled neutron scattering experiments as described below can be groundbreaking.

\section*{Integration of machine learning into neutron scattering}\label{sec:int_ml}

The integration of machine learning into neutron scattering can be resolved into four aspects \cite{Chen_2021}. 
These are I) scattering experiment design and optimization; II) parameter space exploration and information compression; III) structure or property predictions; and IV) parameter space predictions. Figure \ref{fig02_MLoverview} (a) schematically shows their functional relationship and integration. 

ML is necessary to achieve the tasks, and dimensionality reduction/ information compression is a central concept of our design. A NLAE architecture was used to compress structure factor information into a latent space(\latent) of reduced dimension.  The latent space forms the backbone of the operation, into which experimental data, simulations, and predictions from the generative model (GM) feed and from which structure, property, and model parameters are predicted (see Fig. \ref{fig02_MLoverview}).

\subsubsection*{ I) Scattering experiment design and optimization}\label{sec_sub_sub:ml_scattering}
 
Effective design of an experiment involves setting up the instrument and measurement parameters to collect meaningful data to test underlying hypotheses or refine models. The initial hypothesis or model is used to determine the initial instrument parameters and experimental conditions.  As measurements evolve there is a double feed-back process: On one side, processed results at different stages are fed back that change the instrument parameters (stage III to I, see Fig. \ref{fig02_MLoverview} (a)), on the other, improved modeling and predictions feed back into the initial hypothesis, and subsequently into instrument and experiment parameters (stage IV to I).  The latter requires a detailed analysis of the results and is usually unachievable in real time during experiments.  Here ML can make a qualitative change.    

Combining hierarchical clustering using the latent space of the NLAE and the generative model allows hypothetical phase diagrams for the behavior of the material to be constructed. Analysis of the data processed through the autoencoder and its comparison with the measurement allow for differences with the model to be detected. Meanwhile, the variance of the values in latent space determine the degree to which distinguishing features are detected, giving a criterion for sufficiency of counting and measurement. Data sets at other conditions such as field, pressure, and temperature then provide validation. A pre-trained NLAE and generative model speed up the process to the point where feed-back from fully processed results can also be used live in the experiment.

In the case of DTO, the pre-trained NLAE and GM allowed to predict a phase diagram (Fig. \ref{fig01_genModel}(b)) at finite temperature which shows finely balanced structures controlled by further neighbor exchanges. In combination with previously considered phase effects \cite{Samarakoon:2020aa}, trends anticipated from physical variables such as uniaxial and hydrostatic pressure, applied magnetic field, and doping effects can be projected out from such phase diagrams in terms of their control over phase stability, Fig \ref{fig01_genModel}(d). Hypotheses can then be constructed and targeted. 

Here we hypothesize that the morphology of the recently proposed structural glass state \cite{samarakoon2021structural} and its related anomalous noise \cite{samarakoon2021anomalous} in spin ice should be tunable. This would provide a testbed for out-of-equilibrium behavior in a model magnet; an experimental scenario that would facilitate systematic study of long-standing questions regarding breakdown of ergodicity. As indicated in Fig. \ref{fig01_genModel}(d) applied field, doping, and uniaxial pressure are not expected to be effective wherease hydrostatic pressure is expected to appropriately couple into the frustration between interactions. On this basis hydrostatic pressure was selected to tune DTO between phases, combined with temperature to map the development of irreversibility, and the results where analysed in real time to vary experimental parameters.  

\subsubsection*{II) Information compression and parameter space exploration}\label{sec_sub_sub:ml_parameters}

The structure factor determined by neutron experiment, \Sexp, is the observed diffracted intensity at a given scattering vector $Q$ and contains detailed information about the system including correlations and the possible existence of long- and short-range-ordered structures.  Traditionally, direct inspection of \Sexp, and comparison with simulated structure factors, \Ssim, were the tools used for extracting  experimental information. 

With Machine Learning both these processes can be greatly optimised. First, \Sexp, and \Ssim, can be considered volumetric images that need to be analysed and compared, and the usual processes of compression and comparison in latent space used.  Second, the expensive process of calculating structure factors \Ssim, based on model parameters, usually performed using Monte Carlo simulations (see Methods) can be tackled in a more efficient way by means of a surrogate that directly generates compressed representations of model information.

The first part requires an encoder.  Large volumes of $S(Q)$, about $10^6-10^8$ pixels, covering several Brillouin zones, have to be considered. While similar structure factors are expected to indicate phase information of the model and data, the vast $Q$-space dimensionality renders any analysis impractical. To address this dimensionality reduction techniques are needed, {\it e.g.} NLAE or PCA, that compress information to a latent space of much lower dimension ($d_L$), of the order of $10^0-10^2$, while preserving a one-to-one correspondence between the compressed $S(L)$ and the original $S(Q)$. 

For this work we trained  a NLAE architecture comprising an {\it Encoder} and a {\it Decoder} (see methods). The {\it Encoder} takes a linearized version of $S(Q)$ and compress it into the lower-dimensional representation, $S(L)$.  The {\it Decoder} outputs a predicted structure factor, $S_{\rm AE}(Q)$ for any $S(L)$.  The full NLAE architecture is used for the training of the autoencoder itself, minimizing the deviation between the input, a series of simulated structure factors \Ssim, and the filtered output.  

The second part consists in building  a generative network model (GN), a surrogate to bypass computationally expensive direct solvers.  The GN maps model parameter space, $\lbrace H(p)\rbrace$ directly into $S(L)$.   This makes exhaustive searches possible and enables live experiment planning and parameter space mapping.  These predictions depend on the degree of training of the network, and on the topography of the phase space and the sparsity of the sampling.  They do not fully replace simulations and should not be used to draw conclusions when detailed information is needed.  These surrogates can also be used as the low-cost estimator in the iterative mapping algorithm workflow as an alternative to the Gaussian Process Regression as suggested in Ref. \onlinecite{Samarakoon:2020aa}.

Figure \ref{fig01_genModel}(e) shows the design of the surrogate implemented in this work.  A Radial Basis Network (RBN), labeled as {\it Generator} is trained using Monte Carlo simulated data \Ssim (see Methods for details).   A quantitative measure of the predictive power of the GN is given in Figure S2 of the Sup. Material, where a comparison is shown between the predictions for three vectors in \latent given by the GN (solid lines) and those calculated from \Ssim (symbols), as one of the model parameters is varied. The corresponding \Ssim, were not included in the training of the GN or the NLAE.  In the case exemplified, each MC simulated point takes of order one hour to run while the surrogate model takes around a second.

The information from experiments and model (both from simulations and GN) and compressed into \latent is then transferred to processes III and IV.

\subsubsection*{III) Structure or property predictions}\label{sec_sub_sub:ml_surrogate}

In the integration of ML into the scattering pipeline the information encoded into \latent has to be decoded in order to provide some direct feedback into experimental planning (I), and to allow for human-readable comparison between experiments and modeling.  

The decoder section of the NLAE trained previously serves this purpose.  An example of human-readable data comparison is given in Figure \ref{fig01_genModel}(c).  Here the MC calculated scattering patterns (Simulation) \Ssim for five different points in parameter space are compared with the corresponding predictions from the GN model (Surrogate).  The five parameter points are scattered along the  $J_3 - J_{3'}$ plane as indicated by the circled numbers shown in Fig. \ref{fig01_genModel}(b).  These points in parameter space are not part of the training set of the GM.  

The comparison shows that  the predictions are fairly good in most of the regions considered. They are expected to be worse in regions where parameter space is sparsely sampled and the $S(Q)$ is rapidly changing. For example, in the regions where phase transitions or rapid cross-overs occur, as is the case for the point labeled 2.  In this case, more samples are needed in order to increase the prediction accuracy of the surrogate. This is a dynamical process and the surrogate could be  retrained on demand. Even though the prediction accuracy may be weak over some regions, the surrogates are still useful to locate regions with certain correlations that can be later verified with  calculations using more time-demanding simulations.

\subsubsection*{ IV)  Parameter space predictions}\label{sec_sub_sub:ml_categories}

The last section of the ML pipeline also takes as input the latent space \latent.   There are two main tasks within this section: the {\it Latent Space Optimization} or solution of the inverse scattering, which provides feedback into the Model/Hypothesis of section I, and the data {\em Auto-classification} in order to  generate a phase diagram of the system.

{\it Latent Space Optimization}:

The inverse scattering problem is usually an ill-posed one where  ML optimisation can make an important contribution. A ML assisted scheme, illustrated in Fig. \ref{fig02_MLoverview} (b), was recently introduced \cite{Samarakoon:2020aa} that by working in the compressed \latent and introducing a new error measure, $\chi^2_L$ , defined as the sum of the squared distance between latent space vectors of experimental and simulated data, can greatly improve optimization over traditional methods. 

In this work, a variant of the Efficient Global Optimization algorithm as described in Ref. \onlinecite{Samarakoon:2020aa} was employed to minimize $\chi^2_L$  and to map the corresponding uncertainty in the 4-dimensional parameter space of $J_1$ , $J_2$, $J_3$ and $J_{3^{\prime}}$. Figure \ref{fig02_MLoverview}(c) shows the flow-diagram of the iterative optimization mapping algorithm (IMA). A  NLAE of $d_L =30$ was trained to compress data and all the $S^{\rm sim}(L)$ data was used to train the GN to predict $S_{\rm GN}(L)$ for parameters of the Hamiltonian not yet sampled. Unlike Ref. \onlinecite{Samarakoon:2020aa}, here we have calculated a low cost estimator of $\chi^2_L$, denoted by  $\hat{\chi}^2_L$, using the GN rather than the Gaussian Process Regression (GPR). The GN acts also as a denoiser, effectively “averaging out” uncorrelated stochastic errors in the $\chi^2_L$ data (see e.g. Fig. S2). The IMA iteratively collect samples subjected to the condition that $\hat{\chi}^2_L$ is below  the error tolerance threshold, $C_L$, and the GN is retrained.  As more data is collected, the prediction accuracy of the GN towards minimum of the  $\chi^2_L$ becomes higher.

{\it Autoclassification and Phase Diagram Generation}: 

One of the aims of a scattering experiment is to determine a phase diagram for the system, a map of the different types of order the system displays under different circumstances.  The correlations of the system are encoded in $S(Q)$ (and consequently in $S(L)$) and parameter sets corresponding to the same structure would cluster together in  either $Q$ or $L$ space. Thus, archetypal hierarchical clustering can be easily used to classify the main phases. However, such clustering analysis would fail when the system  undergoes continuous transitions or crossovers rather than abrupt first-order like changes.  
In this case it is still possible to easily construct a graphical phase diagram. If the  $Q$ space can be reduced to an $L$ space with $d_L=3$, then the latent vectors can be treated as the RGB color components of a phase map (see ref. \onlinecite{Tennant-2021}). Alternatively, \latent can be further reduced by an additional NLAE with an output into a three dimensional latent space and the resulting vectors treated in the same fashion. 

The case studied here belongs to the second category, since $d_L=30$ after the first reduction.  It was further reduced by a second NLAE, into $L_{c}$ with $d_{Lc}=3$. A separate  GN was trained with the $\lbrace p \rbrace$ and $S(L_c)$ as the input and the target respectively. Figure \ref{fig01_genModel}(b) shows a map of the magnetic orderings (phase map) by varying $J_3$ and $J_{3^{'}}$ , generated in such a manner. 

%
\begin{figure*}
\centering\includegraphics[width=0.99\textwidth]{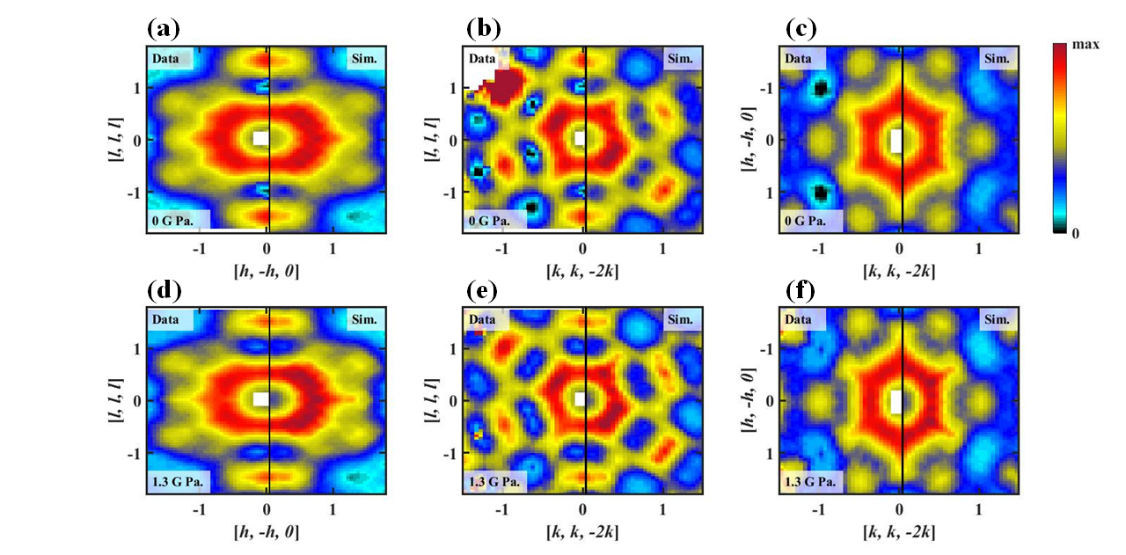}
	\caption{{\em Comparison of magnetic structure factor from both experiment and theory at different pressure}. Three perpendicular slices of 3D volumes of \Sexp are shown for 0 GPa (top) and for 1.3 GPa (bottom). Each panel is a side-by-side comparison of experiment (left) and  simulation (right). All data was collected at $T= 0.68$ K.}
\label{fig03_bestSol}
\end{figure*}
%
%

\section*{Neutron scattering of ${\rm Dy}_2{\rm Ti}_2 {\rm O}_7$ under pressure}\label{sec:results}

An  isotopically  enriched  single  crystal  sample  of Dy$_2$Ti$_2$O$_7$ was used to perform neutron diffuse scattering experiments at the Elastic Diffuse Scattering Spectrometer, CORELLI at the Spallation  Neutron  Source,  Oak  Ridge  National  Laboratory under zero and finite pressure conditions (see Methods for details). 

Figure \ref{fig03_bestSol} shows the magnetic structure factor for three perpendicular slices in reciprocal space at 0 GPa (a,b,c) and 1.3 GPa (d,e,f), all taken at a temperature $T=680$~mK.  The temperature has been chosen so that it is low enough for correlations to be well developed but
sufficiently high to reach equilibrium over a short time scale (see Ref. \onlinecite{Samarakoon:2020aa}). Each panel shows a comparison between the experimental data (left side) and the simulated patterns using the ML optimised parameters (right). In all cases there is a very good agreement between experiments and simulation.  A comparison between the 0 GPa and 1.3 GPa data shows minor changes, corresponding to a slight sharpening with pressure of features already present at 0 GPa, but the structures are qualitatively identical.  Pressure enhances short-range correlations, but does not induce long-range order.  
%
\begin{figure*}
\centering\includegraphics[width=0.99\textwidth]{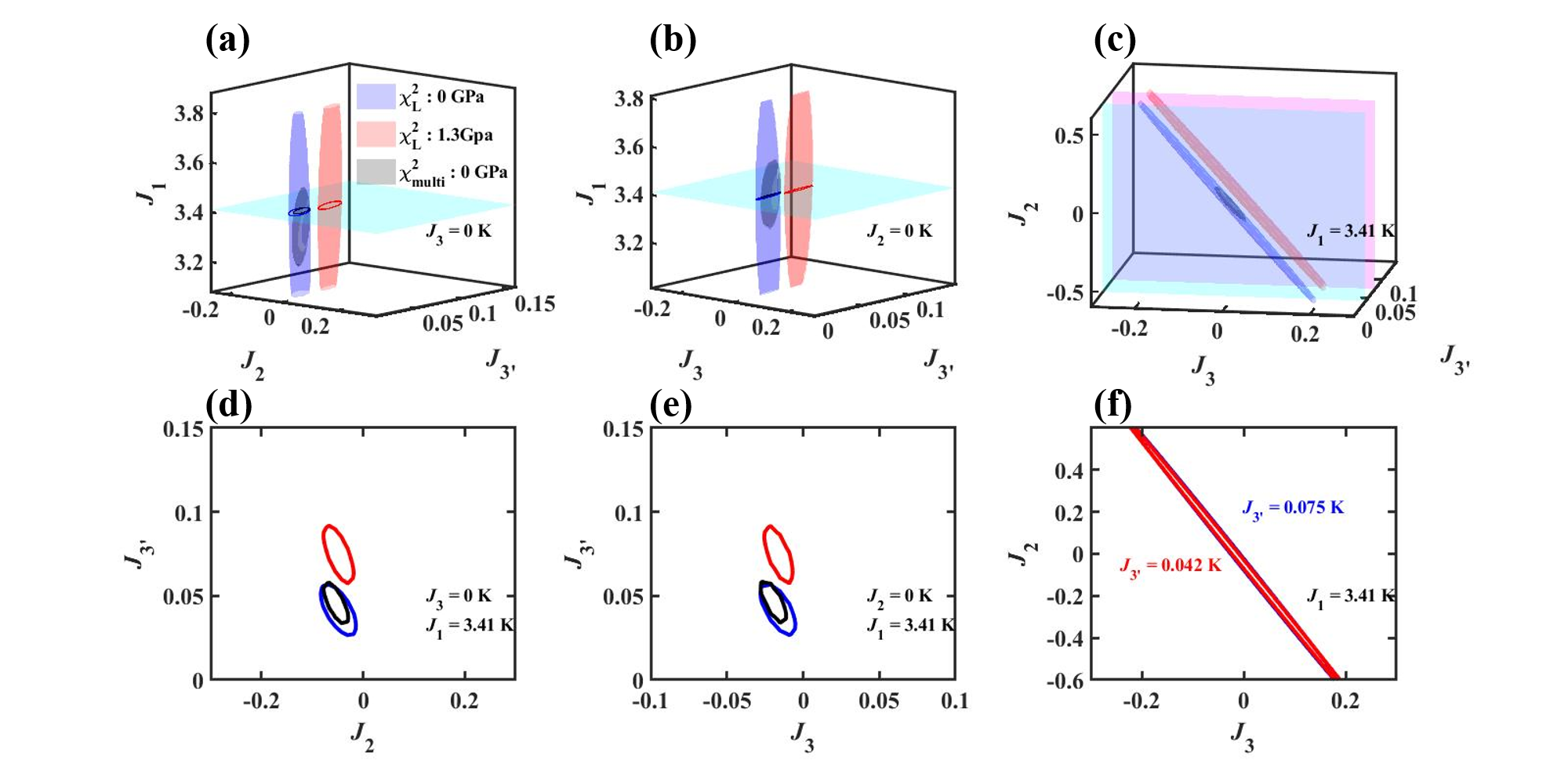}
	\caption{{\em Multi-dimensional solution manifolds of experimental data}. Three-dimensional slices of 4D solution manifolds with (a) $J_3$=0 K, (b) $J_2$=0 K,  (c)  $J_1$=3.41 K.  Two-dimensional slices with (d)  $J_3$=0 K and $J_1$=3.41 K,  (e) $J_2$=0 K and $J_1$=3.41 K , and (f) $J_1$=3.41 K with $J_{3}^\prime$=0.042 and 0.075 K.  The coloured contours denote the the region in parameter space that best fit \Sexp at 0 GPa (blue) and 1.3 GPa (red).  Black contours denote the combined uncertainty for the fit to \Sexp and the heat capacity ($C_v$) at 0 GPa. Both $J_1$ and $3J_2-J_3$ are ill-constrained by \Sexp, yet the relation  $J_2+3J_3=-0.0394 K$ holds for both datasets (0 GPa and 1.3 GPa). Combining \Sexp with $C_v$ further reduces the uncertainties of the solution. Unfortunately, it is extremely hard to measure heat-capacity of a material under pressure. However, the uncertainty along $J_{3}^\prime$ is low enough to clearly resolve a difference of $\approx 3$ mK in $J_{3}^\prime$ as a hydrostatic pressure of 1.3 GPa is applied.	}
\label{fig04_optmalRegion}
\end{figure*}

The ML processing allows for a quick and effective determination of the variation in the Hamiltonian parameters. The variation with pressure in the lattice parameter is negligible up to 1.3 GPa, as determined from the nuclear Bragg peaks.  This means that no variation in the dipolar interaction parameter $D$ needs to be considered, and the parameter space becomes effectively four-dimensional. Figure \ref{fig04_optmalRegion} shows different two and three dimensional cuts of this four dimensional parameter space.    The light blue and red volumes correspond to the region in parameter space where the $\chi_L^2$ is minimised for 0 GPa and 1.3 GPa respectively,  as determined from the comparison with the experimental $S(Q)$. In the case of ambient pressure, this volume can be further reduced by considering also specific heat data (dark blue region).  Panels d) and e) show the outline of the minimum $\chi_L^2$ volumes in the plane $J_1 =3.41$ K (indicated in cyan in panels a) and b)).  This is the optimal value for $J_1$ determined in the literature for 0 GPa by means of a combination of experimental probes \cite{borzi2016intermediate,yavors2008dy,henelius2016refrustration,giblin2018}. While $S(Q)$ leaves ill constrained both $J_1$ and the relation $3J_2 - J_3$, it is clear that there is no overlap between the optimal $\chi^2_L$ volumes, and that the effect of pressure is to induce a  shift in the value of $J_3^\prime$ of $\approx 3$ mK. The optimal parameters for 1.3  GPa are marked in the $J_3 -J_3^\prime$ plane  in Fig.~\ref{fig01_genModel} b). Pressure moves the system deeper into the blue region, where only short-range correlations arising from subsets of ice-states are present.

\begin{figure*}
\centering\includegraphics[width=0.99\textwidth]{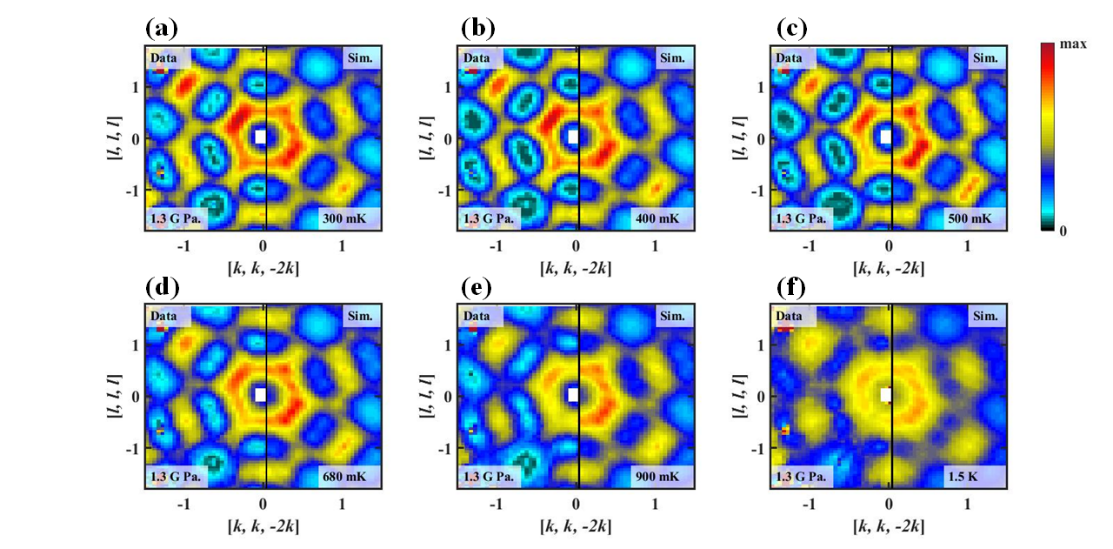}
	\caption{{\em Validation over temperature dependence of scattering data.} Two-dimensional slices of $[l,l,l]-[k,k,-2k]$ at six temperatures: (a) 300 mK, (b) 400 mK, (c) 500 mK, (d) 680 mK, (e) 900 mK and (f) 1.5 K. Each panel is a side by side comparison of experiment (left) and  simulation (right). All the experimental data shown here are collected at 1.3 GPa.  The parameters for the simulations are $J_1$ = 3.3(3) K, $J_2$ = -0.079(8) K, $J_3$ = 0.010(4) K, $J_3\prime$ = 0.075(5) K and $D$ = 1.3224(1) K. }
\label{fig05_validationTDep}
\end{figure*}

The parameters determined for 1.3 GPa can be easily validated by looking at the temperature dependence of $S(Q)$.  Fig.~\ref{fig05_validationTDep} shows a 2D slice of S(Q) in $[l,l,l]-[k,k,-2k]$ for six temperatures (between 300 mK and 1.5 K).  Each panel is a side by side comparison of the experiment, on the left-hand side of the panel, and the simulation, on the right-hand side, using the parameters determined with the ML optimisation for $T=680 mK$.  The agreement at all temperatures between model and experiment is very satisfactory. 

\begin{figure*}
\centering\includegraphics[width=0.99\textwidth]{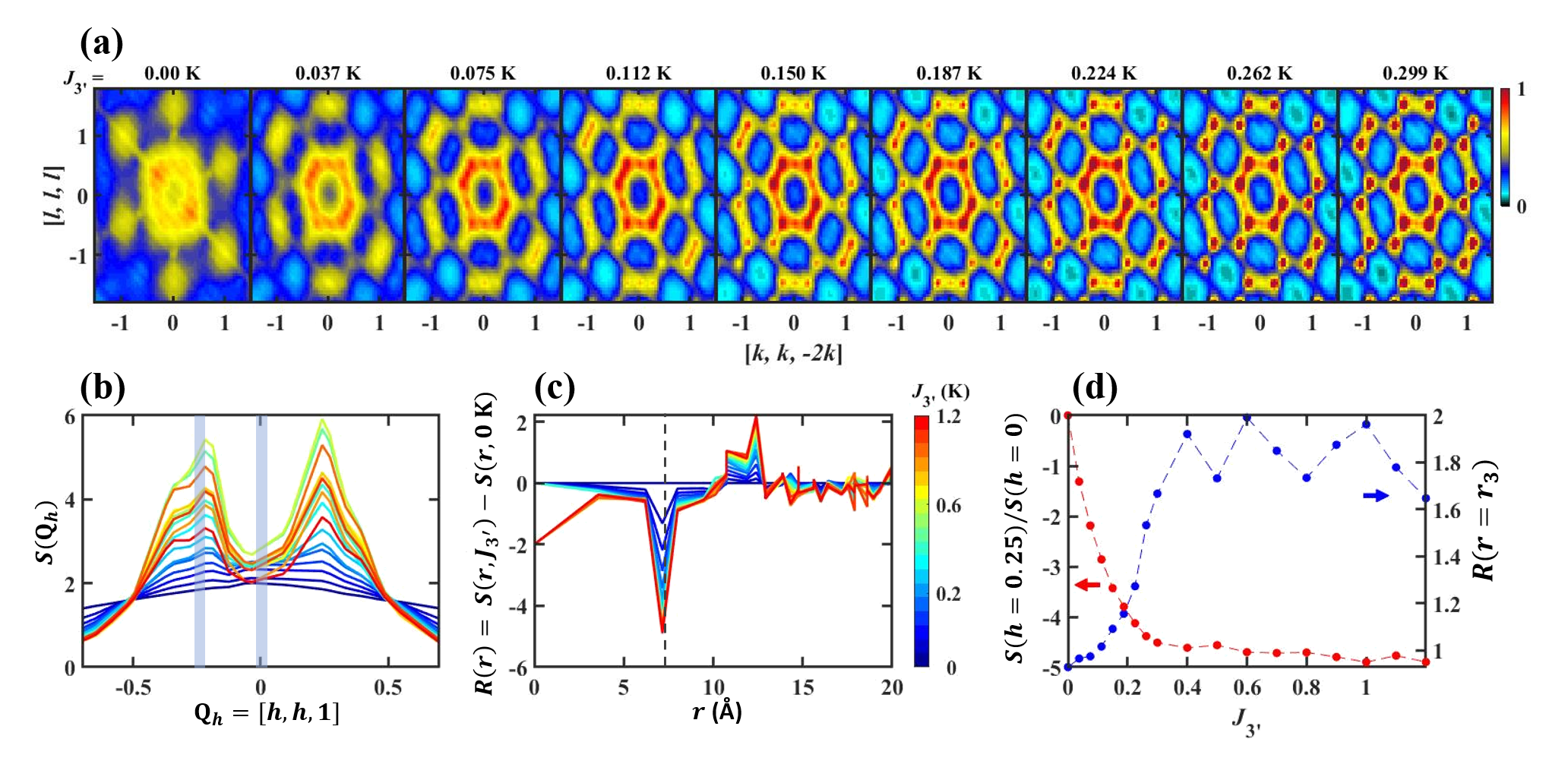}
	\caption{{\em Structural change of the pyrochlore spin-system as a function of $J_3\prime$.} (a) The evolution of calculated \Ssim as a function of $J_3\prime$ with fixed values of other exchange parameters at $J_1$=3.33 K, $J_2$=-0.05 K, $J_3$=0 K and $D$=1.3224 K, (b) line-cuts along $[h,h,1]$, (c) relative spin-spin correlation function, $R(r) = S(r,J_{3'})-S(r, 0 $ K$)$, where $S(r)=\langle S_i^\alpha.S_j^\alpha \rangle$ and (d) comparison of a intensity ratio, $S(h=0.25)/S(h=0)$  and the relative 3rd NN correlation, $S(r_3)$ as a function of $J_3\prime$ (the $Q$-integration ranges are marked as grey areas in panel (b)). 
	}
\label{fig06_Study}
\end{figure*}

The effective Hamiltonian obtained using ML has predictive power and can be used to explore the consequence of a further increase in $J_3^\prime$. Fig.~\ref{fig06_Study} a) shows the evolution of a cut in $[l,l,l]-[k,k,-2k]$ of $S(Q)$ as  $J_3^\prime$ is gradually increased from 0 towards 0.3 K with the other parameters fixed. The features sharpen, but no proper long range order (LRO) is established.  This can be also clearly appreciated in the line-cuts along [h,h,1] plotted in Fig. \ref{fig06_Study} (b).  Even if $J_3$ is increased more than an order of magnitude beyond what is achieved by the application of 1.3 GPa, there is still no development of LRO. This can be clearly seen in the calculated relative spin-spin correlation function, 
$$
S(r,J_3^\prime ) = \sum_{r_{ij}} \mathbf{S}_i\cdot  \mathbf{S}_j
$$
shown in Fig. \ref{fig06_Study} (c), which exhibits a gradual increase in short range spin correlations as $J_3\prime$ is increased, also evident in the relative correlation function (see Fig. \ref{fig06_Study} (d)).

%

%
%
%


\section*{Discussion}\label{sec:discussion}

\subsection*{Machine Learning}

The results here show that the application of machine learning in the planning, collection, analysis, and 
interpretation of neutron experiments and data provides a powerful capability. It is easy to imagine the 
extension of this approach to inelastic data and other codes.  

{\it Instrumental effects}: The effects of instrumental resolution are straightforward in the case of diffuse scattering where 
the signal is broadened and so is not strongly affected. In other applications resolution will be more 
challenging. Surrogates based on Monte Carlo ray tracing simulations of instruments should be able to provide fast capabilities for 
computation of resolution effects. Codes are available, e.g. McSTAS \cite{lefmann1999mcstas} and  McVINE \cite{lin2016mcvine}, by which simulations over 
ranges of cases can be made available for training.

{\it Data Compression}: Non-linear autoencoders are successful at efficiently compressing input from diffuse scattering. 
Other architectures could provide even more effective training. For example Euclidean Neural Networks (ENN)  can exploit the symmetries 
in crystalline samples which may make them trainable with far fewer simulations \cite{PhysRevResearch.3.L012002}. Certainly the relatively small 
dimensionality of the latent space here implies a high degree of compression which a network that encodes some 
physics into may very effectively learn for complex cases. These also open the way for more generic training 
over a wide range of Hamiltonians as well as experimental data itself. Such an ENN could form the basis of a 
more model agnostic compression of experimental data. Combined with Bayesian analysis this would provide a basis 
for assessing experimental data for more autonomous steering of experiments. 

{\it Data processing} A well trained NLAE, as utilized here, successfully undertakes filtering of experimental background and artifacts. Alternative approaches, which are suitable for cases where less precise models are available, involve identifying background and artifacts more straightforwardly. Generally these will provide signals that do not correspond to the underlying symmetries and physical constraints so physics informed networks such as ENNs should provide good discrimination. Another approach is to use both measured background data sets and simulations to train machine learning to identify and remove these signals. Generally, instrumental backgrounds correspond to a limited number of processes such as scattering from sample environments or phonons from sample mountings. 

{\it Integrated workflows}: The simulations and training of the machine learning components as well as data processing requires coordination over data and computing resources. Current instrument control systems at the Spallation Neutron Source and High Flux Isotope Reactor at Oak Ridge National Laboratory are able to integrate feedback from analysis. However, the scheduling and management of the simulations and data processing using machine learning requires an infrastructure that can combine edge and high-performance computing resources. This requires to work over a delocalized network where specialized codes will be required to run at other institutions. These developments will be needed if machine learning is to transform experiments beyond simpler cases of diffraction and small angle scattering where large data bases and standard codes are available.

{\it Modeling}: A large number of approaches to the simulation of magnetic structure and dynamics are available. Our work for elastic and inelastic scattering has utilized Monte Carlo and Landau Lifshitz approaches which cover a wide range of cases \cite{Tennant-2021}. Spinwave theory provides fast computation for simple magnon dynamics and codes are available \cite{Tth2015LinearSW}. More interestingly, machine learning holds the promise of being able to interface sophisticated simulations \cite{Butler_2021} that include quantum effects including correlations which are important to quantum materials. Examples are density matrix renormalization group \cite{RevModPhys.77.259}, quantum Monte Carlo \cite{Zhang2004}, dynamical mean field theory \cite{RevModPhys.78.865} and dynamic cluster approximation approaches to name but a few. Efficient training is of paramount importance as these are computationally expensive methods to run. However, bringing such state-of-the-art theoretical methods closer to experiment would undoubtedly have a significant impact on our understanding of quantum materials. Further, artificial intelligence is being used to accelerate these methods and can be expected to facilitate their integration with experiment in the foreseeable future.

{\it Applications}: The general approach to machine learning integrating into neutron scattering is potentially applicable to a wide range of science cases. A wide array of diffuse scattering problems could be approached this way with large scale atomic simulations being an obvious starting point. Inelastic data is well suited also. Crystal field measurements should be considered and may well be open to a significant degree of automation of measurements and analysis. Powder inelastic scattering is also well suited which could make a significant impact on throughput and time to understand materials. Finally, single crystal inelastic scattering from quantum magnets, itinerant and superconducting materials, and anharmonic phonons are obvious targets.

\subsection*{Pressure control of a magnetic glass}

Our results show that the further-neighbor couplings are successfully tuned by hydrostatic pressure. 
The pressure applied of $\approx 1.3$~GPa perturbs the system modifying the nanoscale magnetic order from ambient pressure.
The morphology of the resulting glass \cite{samarakoon2021structural} restricts monopole pathways and is a key property reflecting the phase ordering kinetics.
Recently, we have proposed spin ices as a model systems to explore glass formation systematically \cite{samarakoon2021structural,samarakoon2021anomalous}. The results here are promising; diffuse neutron studies reaching $\sim 7$~GPa are feasible. Over such a range significant variation would be attainable and the role of subtle interactions on the development of non-equilibrium phases rigorously testable.
Anomalous dynamics, such as colored noise spectra, are signatures of memory effects \citep{samarakoon2021anomalous} and when combined with diffuse scattering give a comprehensive characterization of the state. The systematics then could provide a powerful connection between theory and experiment for the long standing and difficult problems involving the fundamentals of glass and correlated liquid behavior; a connection made possible by ML based approaches.

\section*{Summary and Conclusion}
In this paper we have proposed a scheme for the application of machine learning to neutron scattering. The approach uses non-linear autoencoders to undertake compression to a latent space from training data involving computationally expensive simulations of neutron scattering data. A generative model based on the NLAE provides fast calculations which allow identification of areas of interest and experiment planning. Hierarchical clustering provides categorization of theoretical phase diagrams and identification of experimental phases from measurements. The NLAE also provides capabilities for accurate parameter determination and data treatment/handling. We explore these capabilities on the highly frustrated magnet Dy$_2$Ti$_2$O$_7$ under pressure. This material has a complex physical behavior which can be extracted rapidly from the combined measurements and data analytics based on simulations. Our analysis shows that hydrostatic pressures of up to $1.3$~GPa are able to modify the magnetic interactions of the material leading to the prediction that substantially higher pressures may cause a magnetic phase transition. This provides a novel route to a pressure tunable structural glass. 

%
\section{Acknowledgements}
We acknowledge useful discussions on machine learning and workflows with Mingda Li, Tess Smidt, Scott Klasky, Juan Restrepo, Cristian Batista, and Guannan Zhang; and on glass formation in spin ice with Roderich Moessner and Claudio Castelnovo. A portion of this research used resources at the Spallation Neutron Source. The research by D. A. T. was sponsored by the Quantum Science Center. A.M.S. was supported by the U.S. Department of Energy, Office of Science, Materials Sciences and Engineering Division and Scientific User Facilities Division. S.A.G. acknowledges support from  Agencia Nacional de Promoci\'on Cient\'\i fica y Tecnol\'ogica through PICT 2017-2347. The computer modeling used resources of the Oak Ridge Leadership Computing Facility, which is supported by the Office of Science of the U.S. Department of Energy under contract no. DE-AC05-00OR22725.

\section{Methods}

\subsection{Training of the generative model}

 The RBN is constructed with two layers. A layer of radial basis (RB) neurons followed by an output layer of  logistic neurons. The latent space predictions for a given parameter set $\lbrace p \rbrace$ are defined as: \begin{eqnarray}
\mathcal{L}_i&=& f_2(w^{(2)}_{ij}.h_j(\lbrace p \rbrace) + b_2)
\end{eqnarray}\begin{eqnarray}h_j(\lbrace p \rbrace) = exp\left\lbrack -\frac{\sum_{k} (p_k - c_{jk})^2}{\sigma^2} \right\rbrack\end{eqnarray}where  $f_2$ is the logistic activation  function $f(x) = 1/(1+e^{-x})$  similar to output layer of the NLAE {\it Encoder}. The weight matrix $w^{(2)}_{ij}$, and bias vector $b_2$ of the output layer and the clustering centers $c_{jk}$  of the RB layer  are to be determined from the training process. The spread of RB function, $\sigma$  is preset to 0.05. The network is trained with the $S^{\rm sim}(L)$  as the target and the corresponding  $\lbrace p \rbrace$ as the input. Thus the input and the output dimensionality is set by the dimensionality of the $\lbrace H(p)\rbrace$ and the $\mathcal{L}_Q$. The number of neutrons in the RB layer is determined during the training process. The training starts with no neurons in the hidden layer and iteratively adding neurons to minimize the error between output and the target.  

\subsection{Monte Carlo based Direct Solver for scattering $\mathcal{S}^{\rm sim}(Q;\lbrace p \rbrace)$}\label{sec_sub:direct_solver}

The Monte Carlo based direct solver for the neutron scattering used in Ref. \onlinecite{Samarakoon:2020aa} is used here and we summarize the pertinent details. In DTO the individual vector spins ${\mathbf S}_i=\left[ S_i^x,S_i^y,S_i^z \right]$ are at positions $\mathbf{R}_i$ which are located on the pyrochlore lattice and behave as classical Ising spins that can point in or out of the tetrahedra. The energy due to interactions is given by the spin Hamiltonian, $\mathcal{H}=\mathcal{H}(\lbrace p \rbrace,\mathbf{S},\mathbf{R})$, where $\lbrace p \rbrace$ is the set of interaction parameters. In the case of the dipolar spin-ice Hamiltonian that includes exchange terms up to third-nearest neighbors:
\begin{eqnarray}
\mathcal{H}&=&\sum_{\alpha=1,2,3,3'} J_{\alpha} \sum_{{\lbrace i,j \rbrace}_{\alpha}} 
S_i \cdot S_j + \nonumber \\ 
&&+ \mathcal{D} r_1^3 \sum_{\lbrace i,j \rbrace} 
\left\lbrack \frac{S_i.S_j}{| r_{ij}|^3}
-\frac{3({\bf S}_i.{\bf r}_{ij}  ).({\bf S}_j.{\bf r}_{ij}  )}{| r_{ij} |^5} \right\rbrack .
\end{eqnarray}
The model includes first, second, and two different third nearest neighbors with interaction strengths, $J_1$, $J_2$, $J_3$ and $J_{3'}$ respectively. 
There is also a dipolar interaction with strength $\mathcal{D}$, which couples the $i^{th}$ and the $j^{th}$ spins according to their displacement vector $r_{ij}$. The set of interactions spans $\lbrace p \rbrace=\lbrace J_1,J_2,J_3,J_{3'},\mathcal{D} \rbrace$ and $\mathcal{D} = 1.3224$~K is fixed here.

Realistic spin configurations can be prepared through annealing based on the Metropolis algorithm, a Markov Chain Monte Carlo method \cite{metropolis1953equation}. The Metropolis algorithm anneals the configuration $\mathbf{S}$ to be representative of the system in thermal equilibrium at chosen temperature $T$. From these configurations a full range of physical properties can be calculated.

The diffuse scattering from the magnetic system is (approximately) proportional to the cross section \cite{lovesey1984theory}:
\begin{eqnarray}
\frac{d^2\sigma}{d\Omega}&=&r_m^2 \sum_{\alpha,\beta} \frac{g_{\alpha}g_{\beta}}{4} \left(\delta_{\alpha\beta}-
\frac{q_{\alpha}q_{\beta}}{q^2}\right) \times \nonumber \\
& &\times|F(\textbf{Q})|^2
\mathcal{S}^{\alpha\beta}\left( \textbf{Q}\right)
\label{eq:neutron_scattering}
\end{eqnarray}
where \textbf{Q} is the wavevector transfer in the scattering process, $r_m$ is a scattering factor, $\alpha,\beta=x,y,z$ are cartesian coordinates indicating initial and final spin polarization of the neutron, $F(\textbf{Q})$ is the magnetic form factor and $\mathcal{S}^{\alpha\beta}\left( \textbf{Q} \right)$ is the scattering factor correlation function:
\begin{equation}
    \mathcal{S}^{\alpha\beta}(\textbf{Q})=\frac{1}{2\pi N}\left| S_Q^{\alpha} S_{-Q}^{\beta} \right|
\end{equation}
with 
\begin{equation}
    S_Q^{\alpha}=
 \sum_i S_i^{\alpha} (t_n) e^{i\textbf{Q}\cdot \textbf{R}_i}.
\end{equation}
The actual measured cross sections $\mathcal{S}^{\rm exp}\left( \textbf{Q} \right)$ depend on experimental 
conditions including resolution and the direct solver then undertakes the transformation $\lbrace \mathcal{H}(p) \rbrace \rightarrow \mathcal{S}^{\rm sim}(\textbf{Q},\lbrace p \rbrace)$ to replicate expected scattering.

\subsection{Experiment Details}\label{sec:experiment}

\subsubsection{Crystal Growth}\label{sec_sub:crystal}

For this work we used an isotopically enriched single crystal sample of Dy$_2$Ti$_2$O$_7$, grown using floating-zone mirror furnace (see \onlinecite{Samarakoon:2020aa}). The single crystal was cut and polished to be cylindrical.  The diameter, height, and mass are 1.8 mm, 5 mm, and 76.6 mg respectively. The cylindrical axis of the sample aligns with [h,-h,0] crystallographic direction. The polished crystal was kept inside a Teflon tube filled with Fluorinert FC-770 as a pressure transmission medium, and a Copper-beryllium cell was used to apply hydrostatic pressures. 
A load of ~0.9 tonnes was applied using a hydraulic press, and the resultant pressure at the sample can be estimated as $\sim1.3$~GPa, according to Fig. 4 of ref. \citep{komatsu2015zr}. 

\subsubsection{Diffuse Neutron Scattering Experiment}\label{sec_sub:diffuse_expmt}

Elastic Diffuse Scattering Spectrometer, CORELLI at the Spallation Neutron Source, Oak Ridge National Laboratory was used to perform experiments under zero and finite pressure conditions\cite{ye2018implementation}. Ambient pressure and 1.3 GPa experiments were performed on two different beam-time and experiment setups. The dilution refrigerator insert was used in both cases to enable the measurements down to 100 mK. A cryomagnet was used in the ambient pressure experiment, and two datasets at temperatures of 100 mK and 680 mK were collected under zero-field \cite{Samarakoon:2020aa} In the pressure experiment, the loaded pressure cell was rotated through 360 degrees with the steps of 3 degrees horizontally in the [H,H,L] plane of reflection at a fixed temperature. Seven measurements were repeated at 300 mK, 400 mK, 500 mK, 680 mK, 900 mK, 1.5 K, and 19 K. The data was reduced using Mantid [24] and Python scripts available at Corelli.

%

\onecolumngrid
\newpage
\clearpage
\setcounter{equation}{0}
\setcounter{figure}{0}
\setcounter{section}{0}
\setcounter{page}{1}
\makeatletter 
\renewcommand{\thefigure}{S\arabic{figure}}
\renewcommand{\theequation}{S\arabic{equation}}
\renewcommand{\thepage}{S\arabic{page}}
\setlength\parindent{10pt}

\section*{Supplementary information for: Integration of Machine Learning with Neutron Scattering: Hamiltonian Tuning in Spin Ice with Pressure}

\renewcommand{\labelenumi}{\Roman{enumi}.}

This file includes:
\begin{enumerate}
    \item  Dy$_2$Ti$_2$O$_7$ crystal and the pressure Cell
    \item  Formulation of the optimal region 
    \item  Quantitative predictions of the generative model
    \item  Temperature dependence of ambient pressure solution
    \item  Methods to constrain the value of 
    \item  Additional numerical studies as a function of  
\end{enumerate}
Supplementary Figures 1-5

\section{\dto crystal and pressure cell}

\begin{figure}[H]
\centerline{\includegraphics[width=0.8\columnwidth]{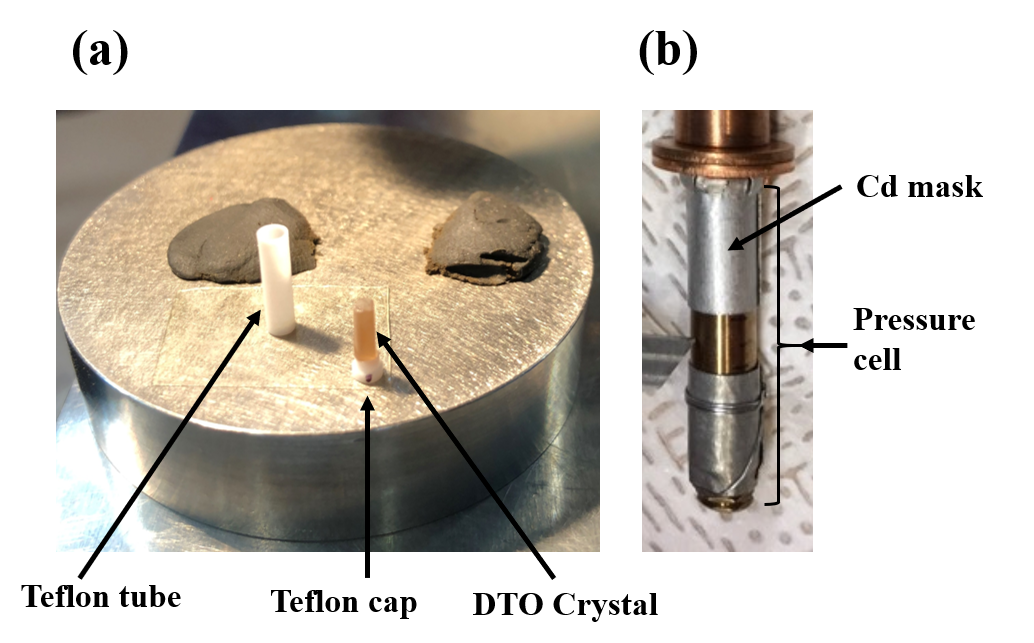}}
	\caption{(a) The Teflon tube and the polished Dy2Ti2O7 single-crystal attached to the Teflon cap. (b) The Copper-beryllium cell with a Cd mask covering outside the Teflon tube window.  }
\label{S1}
\end{figure}

As shown in Figure \ref{S1}(a), a cylindrical single-crystal (SL) of \dto was used for the pressure experiment. First, the sample was polished to fit the Teflon tube and glued to the Teflon cap. Next, the Teflon tube was filled with  Fluorinert FC-770 as a pressure transmission medium before installing the crystal. Then, the installation was done carefully not to trap any air bubbles inside, and a Copper-beryllium cell was used to apply hydrostatic pressures, as shown in Fig. \ref{S1}(b). Finally, the Cadmium was used to mask the cell portions except where the DTO crystal is present.  

\section{Formulation of the optimal region}

The four-dimensional optimal region for which $\chi^2_{\rm L}$ (0 G.Pa) $< C^2_{\rm L}$, as illustrated in Fig. 4 (blue contours), can be formulated by fitting the region to a minimum volume ellipsoid \cite{moshtagh2005minimum} as,
\begin{displaymath}
V \times M \times V^\dagger \leq 1
\end{displaymath}
with
\begin{displaymath}
M =
\begin{bmatrix}
4 & 7 & 10 & 25\\
7 & 1300 & 3910 & 1909\\
20 & 3910 & 11774 & 5829\\
25 & 1909 & 5828 & 6998
\end{bmatrix},
\end{displaymath}
and
\begin{displaymath}
V=[J_1 - 3.336 J_2 - 0.008 J_3 + 0.019 J_3^\prime - 0.042].
\end{displaymath}
For the minimum volume episode for  the condition $\chi^2_{\rm L}$ (1.3  G.Pa) $< C^2_{\rm L}$, as illustrated in Fig. 4 (red contours),
\begin{displaymath}
M =
\begin{bmatrix}
4 & -16 & -47 & -31\\
-16 & 1958 & 5858 & 2255\\
-47 & 5858 & 17536 & 6752\\
-31 & 2255 & 6752 & 5927 
\end{bmatrix},
\end{displaymath}
and
\begin{displaymath}
V=[J_1 - 3.446 J_2 + 0.001 J_3 + 0.016 J_3^\prime - 0.075].
\end{displaymath}
 
The minimum volume episode for the condition  $\chi^2_{\rm multi}$ (0  G.Pa) $< C^2_{\rm multi}$, as illustrated in Fig. 4 (black contours),

\begin{displaymath}
M =
\begin{bmatrix}
41 & -118 & -203 & -158\\
-118 & 2601 & 7328 & 3612\\
-203 & 7328 & 21226 & 10443\\
-158 & 3612 & 10443 & 10180 
\end{bmatrix},
\end{displaymath}
and
\begin{displaymath}
V=[J_1 - 3.274 J_2 + 0.114 J_3 - 0.02 J_3^\prime - 0.045].
\end{displaymath}

\section{Quantitative predictions of the generative model}

\begin{figure}[H]
\centerline{\includegraphics[width=0.4\columnwidth]{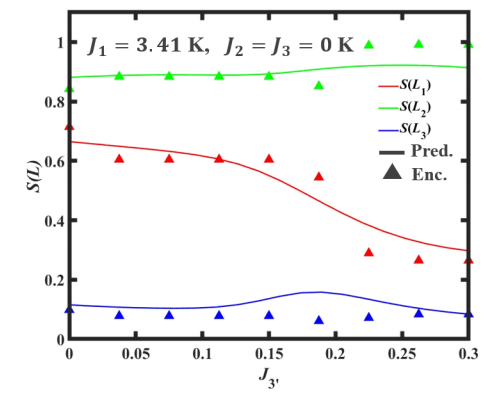}}
	\caption{Comparison between the latent space coordinates, S(L), for the simulated structure factor (filled triangles) and those predicted from the GM (solid lines) for an increasing value of the exchange parameter $J_3^\prime$}
\label{S2}
\end{figure}

 The generative model (GM) can directly calculate the latent space coordinates based on the microscopic parameters of the Hamiltonian, thus bypassing the computer-intensive Monte Carlo Simulations.  Fig. \ref{S2} shows a quantitative comparison of the predictions from the GM compared with direct MC calculations. 

\section{Temperature dependence of ambient pressure solution }

\begin{figure}[H]
\centerline{\includegraphics[width=0.7\columnwidth]{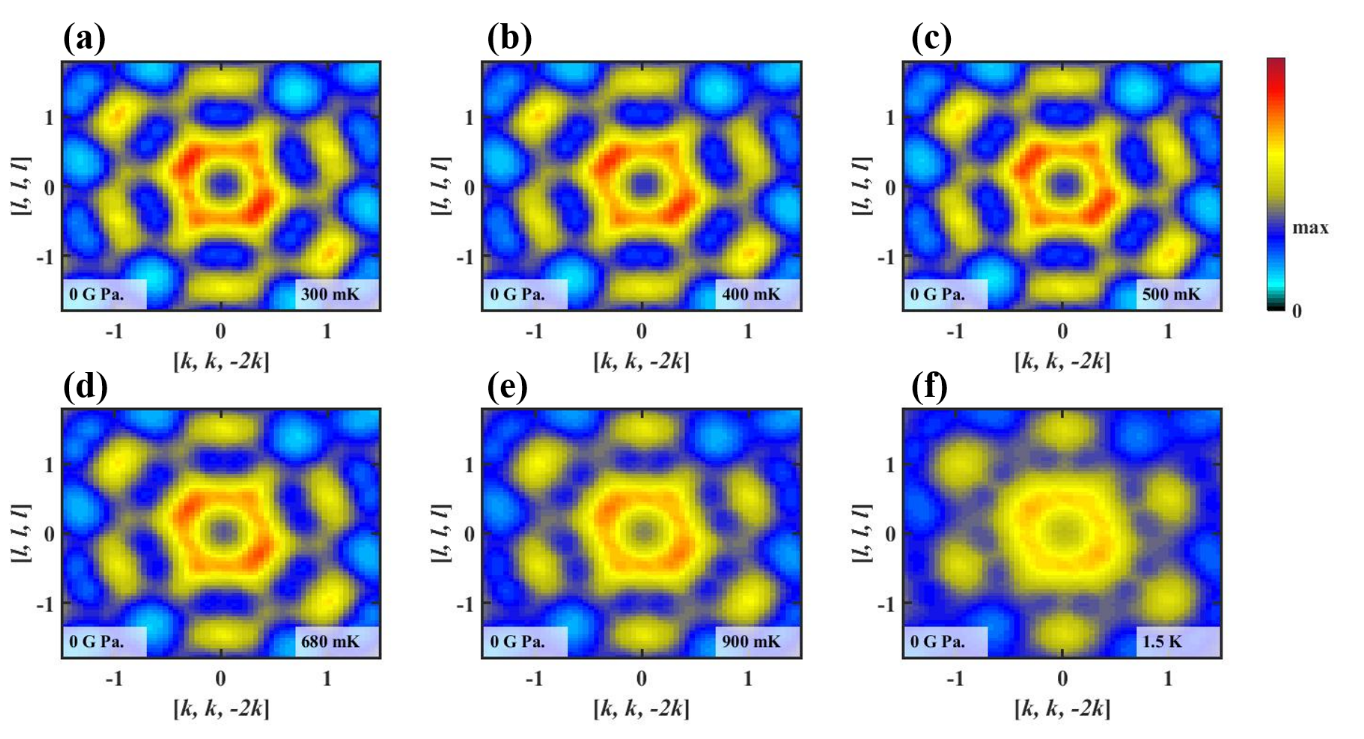}}
	\caption{The simulated structure factors for the optimal Hamiltonian parameters to the 680 mK ambient pressure data set at (a) 300 mK, (b) 400 mK, (c) 500 mK, (d) 680 mK, (e) 900 mK and (f) 1.5 K.}
\label{S3}
\end{figure}

The temperature dependence of  $S(Q)$  for the 0 G.Pa. solution is shown in Fig. \ref{S3}. Similar temperature dependence is shown in Fig. 5 in the main text for the solution at 1.3 GPa. Both solutions show diffuse scattering for Coulombic correlations associated with the isotropic U(1) gauge liquid at 1.5 K.

\section{ Methods to constrain the value of $J_1$}

\begin{figure}[H]
\centerline{\includegraphics[width=0.75\columnwidth]{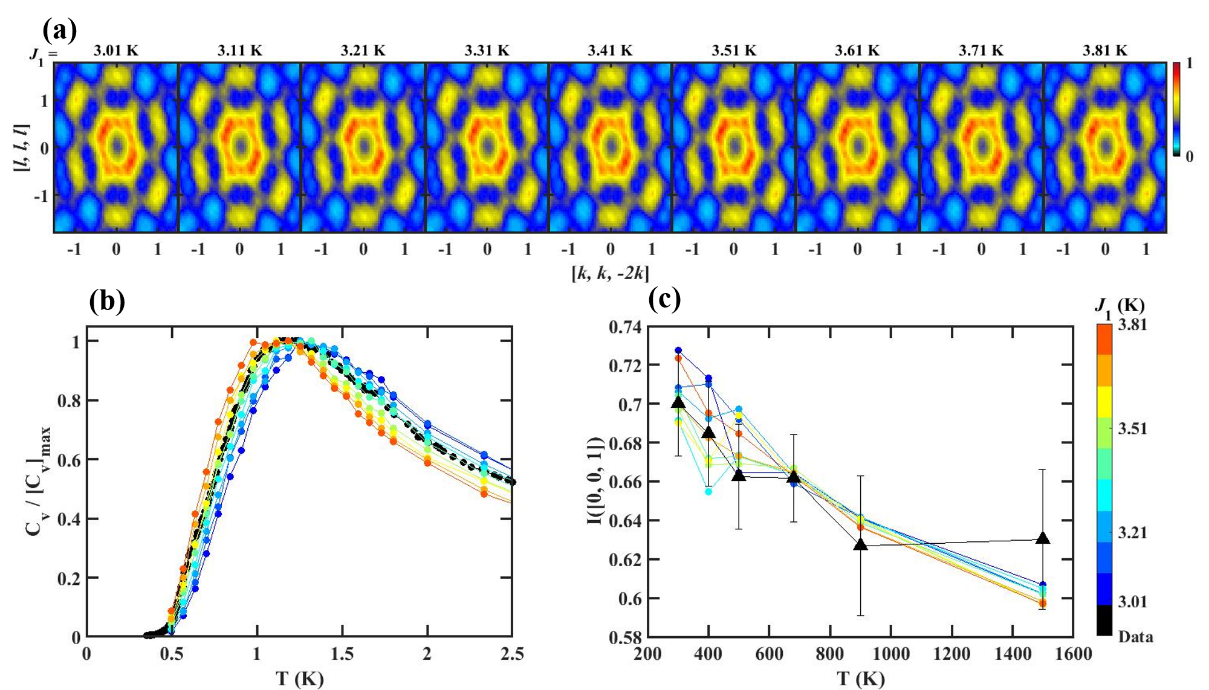}}
	\caption{ (a) Structure factor and (b) heat capacity as a function of $J_1$ at fixed $J_2 = 0.008$ K, $J_3 = -0.019$ K, $J_3^\prime = 0.042$ K and $ D= 1.3224$ K. (c) the integrated intensity of the [0,0,1] diffuse peak as a function of temperature and $J_1$. }
\label{S4}
\end{figure}

The optimal region for just $S(Q)$ does not constrain $J_1$ and  $J_2 - 0.335 J_3 = 0.0144$ K. As shown in Fig. \ref{S4}(a), the $S(Q)$ does not change as a function of $J_1$ at fixed $J_2= 0.008$ K, $J_3 = -0.019$ K, $J_3^\prime = 0.042$ K and $ D = 1.3224$ K. The temperature dependence for each $J_1$  is also checked, and the  $S(Q,T)$ is also not a good discriminator as summarized in Fig. \ref{S4} (c).  However, the heat capacity provides a way to refine $
J_1$ further, as demonstrated in Fig. \ref{S4}(b), and Fig. 4. Only the heat capacity at ambient pressure is available.

\section{ Additional numerical studies as a function of $J_3^\prime$}
 
\begin{figure}[H]
\centerline{\includegraphics[width=0.8\columnwidth]{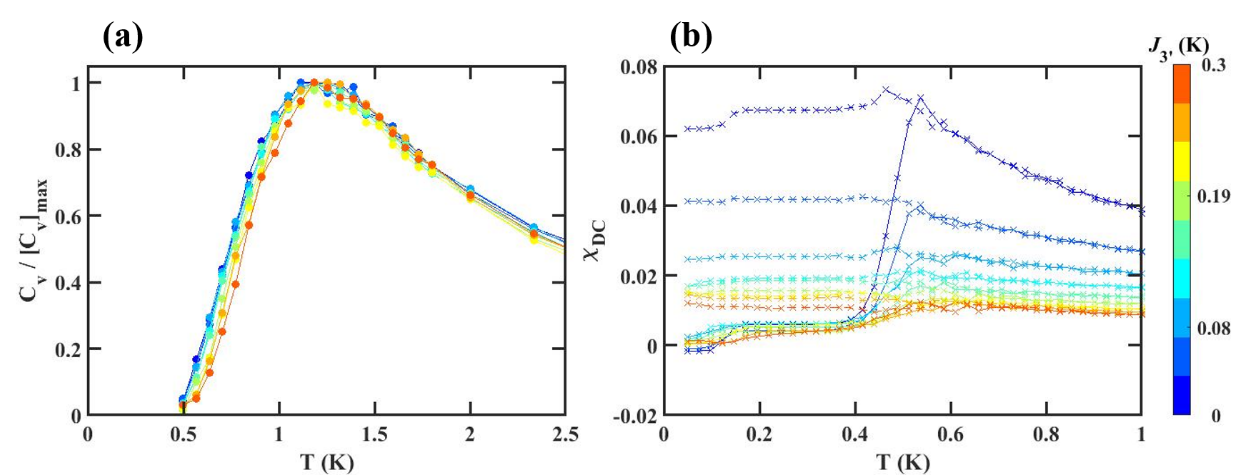}}
	\caption{(a) Simulated Heat capacity and (b) Zero-Field-Cooled (ZFC) and Field-Cooled (FC) susceptibility as a function of temperature by varying $J_3^\prime$ at $J_1 = 3.33$ K, $J_2 = -0.05$ K, $J_3 = 0$ K and $D = 1.3224$ K.}
\label{S5}
\end{figure}

Heat capacity, Zero-Field-Cooled (ZFC) and Field-Cooled (FC) susceptibilities were calculated for the same parameter sets used in Fig. 6. The heat capacity seems to evolve slightly with $J_36\prime$ (Fig. \ref{S5} (a)). However, as discussed in the main text, the model crosses over from one diffuse phase to another, as evident by their correlations (see Fig. 6(d)). Even though the irreversibility temperature, $T_{\rm irr}$
does not any change with $J_3^\prime$, the susceptibility $\chi_{\rm DC}$  seems to have a different temperature dependence below  $T_{\rm irr}$.  This behavior may reflect the local structure of the magnetic phase deep into $J_3^\prime$ is different from the inter-twined ferromagnetic domain structure proposed in ref \cite{samarakoon2021structural}. More numerical investigations are needed to learn the underlying physics of this phase.

\end{document}